\documentclass[aip,rsi,reprint,graphicx]{revtex4-1} 
\usepackage{amsmath,amsfonts,amsthm,bm}
\usepackage{dcolumn}
\usepackage{bm}
\usepackage{graphicx}
\usepackage{subfigure}
\usepackage{color}
\usepackage{scalerel,amssymb}
\usepackage{hyperref}

\draft 

\begin{document}


\title{Simultaneous MOKE imaging and measurement of magneto-resistance with vector magnet: a low noise customized setup for low field magnetic devices and thin films characterization} 



\author{Imtiaz Noor Bhatti}
\email{imtiaz-noor.bhatti@unicaen.fr}
\affiliation{The CNRS, GREYC, ENSICAEN, Normandie University, UNICAEN, Caen-14050, France}
\author{Ilyas Noor Bhatti}
\affiliation{Department of Physics, Jamia Milia Islamia University, New Delhi-110025, India}
\author{L. G. Enger} \author{P. Victor} \author{B. Guillet} \author{M. Lam Chok Sing} \author{O. Rousseau} \author{V. Pierron} \author{S. Lebargy}  \author{J. Camarero} \author{L. Mechin} \author{S. Flament}
\email{stephane.flament@ensicaen.fr}
\affiliation{The CNRS, GREYC, ENSICAEN, Normandie University, UNICAEN, Caen-14050, France}


\date{\today}

\begin{abstract}
Here we report a custom design setup for the simultaneous measurement of magneto-resistance and MOKE imaging in longitudinal configuration for characterization of magnetic thin-films and sensors. The setup is designed so as to cope with a small signal to noise ratio of initial images and to allow sensitive magnetoresistance (MR) measurement at low field. An improved differential algorithm is used to get a good enough contrast of the magnetic images and get rid of beam illumination or small camera gain fluctuations. Home made power supply and pre-amplifier stage were designed so as to reduce the low frequency noise as well as the thermal electrical noise. A vector magnet is used to produce rotational magnetic field so as to study the magnetic anisotropy and calculate the anisotropic constants in magnetic thin films. Magnetoresistive sensors patterned on epitaxial La$_{0.67}$Sr$_{0.33}$MnO$_3$ (LSMO) thin film have been characterized with this setup. The Images of the magnetization reversal process as well as the local magnetization loops deduced from these images provided evidence of a magnetic uniaxial anisotropy induced by the vicinal substrate. The magnetic anistropic constant of the films was then inferred from the MR measurements.


\end{abstract}

\pacs{}

\maketitle 
\section{Introduction}
Magnetic thin films and nanostructures are of prime importance because of their use in devices applications for instance anisotropic magneto-resistance sensors, magnetic logic devices, memory devices etc.\cite{tiginyanu, chopra, nickel, sellmyer, fortin} The magnetic properties like magnetization reversal, anisotropy, magneto-resistance etc are well utilized to realize devices with desire properties.\cite{schafer, belhi, phillips, podder, primus,  boschker, van, ding, kwon} Understanding magnetization reversal, domain wall motion and anisotropic magneto-resistance received much interest in thin films and nanostructures. Besides many other experimental tool available for magnetic characterization of thin films, exploration of magnetic properties and spin reorientation in magnetic thin films are extensively studied using magneto-optical Kerr effect (MOKE). Is is one the different magneto-optical microscopy schemes\cite{McCord}, which depends linearly with the magnetization. This noninvasive method for probing the magnetization reversal in thin magnetic films relies namely on the rotation of the polarization plane of an incident plane polarized light after reflexion on the surface of the magnetized material (see Fig. 1d)\cite{freiser}. 
MOKE measurements can be performed in three possible configurations,  longitudinal, transverse and polar. These three configurations are described by the alignment of magnetization vector with respect to plane of incidence i.e sample plane and plane of light scattering.\cite{yang}. In the longitudinal configuration the magnetization vector lies in the plane of the sample and is in the light scattering plane as well as shown in Fig. 1a. In the transverse configuration the magnetization vector is in the sample plane but perpendicular to the light scattering plane as shown in Fig. 1b. In the polar configuration the magnetization vector is in the light scattering plane but perpendicular to the sample plane Fig. 1c.\cite{stupak}


\begin{figure}
	\centering
		\includegraphics[width=8cm]{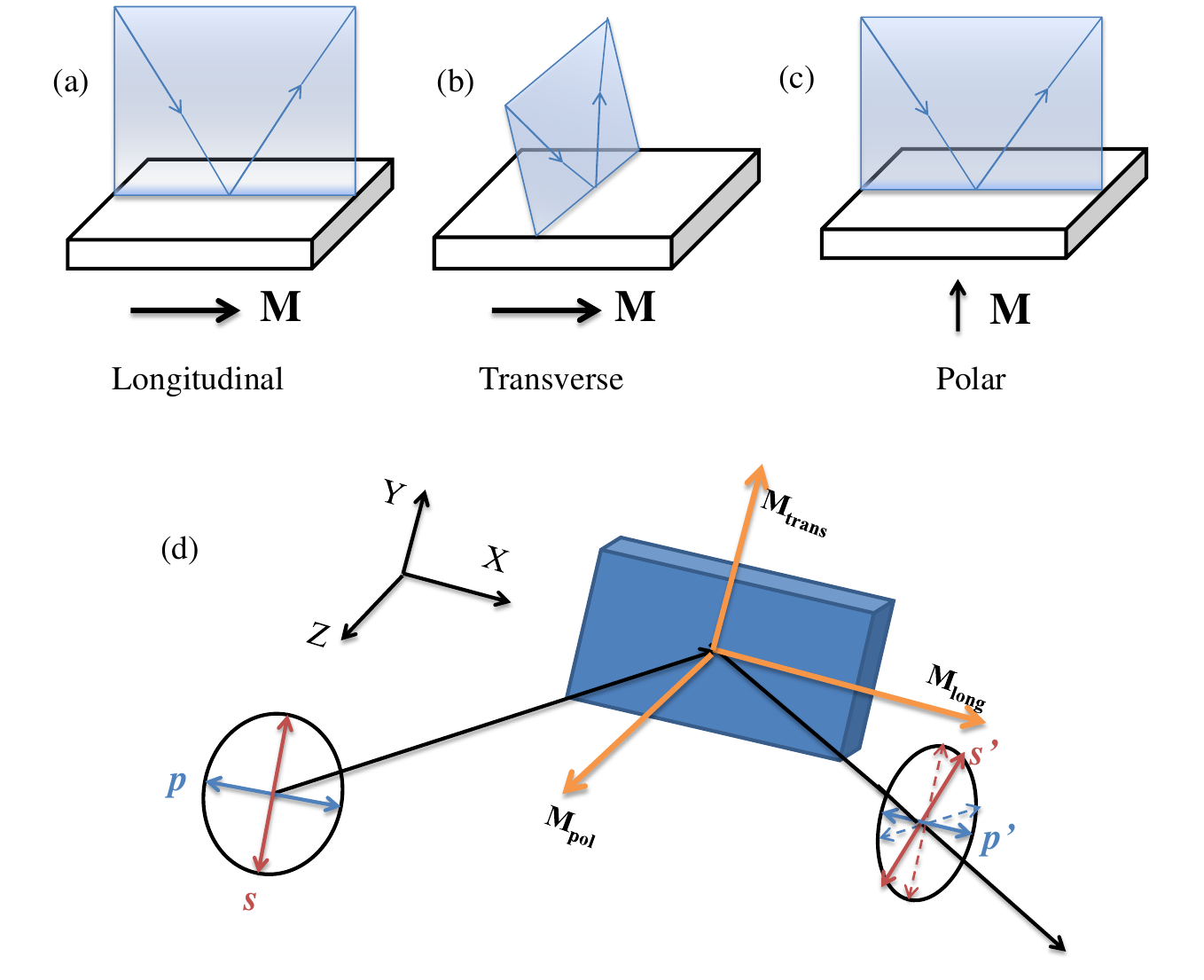}
	\caption{(Color online) Schematic representation of longitudinal, transverse and polar magneto-optic Kerr configuration. Typical representation of magneto-optic Kerr rotation from magnetic sample is shown in lower panel of the Figure.}
	\label{fig:Fig1}
\end{figure}

Many custom instruments based on MOKE have been designed and reported. For instance, vector MOKE is designed to measure the MOKE loop at local area on sample surface\cite{hfding}, similarly polarization modulation technique is employed for vector MOKE magnetometer.\cite{vava} For the study of ultra thin films all three configuration in situ surface MOKE was reported by Lee \textit{et al}.\cite{lee}. Further combine surface stress and MOKE measurement setup was realized to study the stress effect in thin films on magnetic properties.\cite{premper} Direct imaging of domains and domain wall motion at micro scale is also possible with customized designed MOKE.\cite{stupa} For the analysis of near surface magnetic properties of thin films a conversion electron Mossbauer spectrometry and MOKE is combined by Juraszek \textit{et al}.\cite{juraszek}  Combined MOKE loop tracing and magneto-resistance setup is utilized for magnetic anisotropy study in thin films\cite{jli}, 

However when the MOKE signal is differential imaging is often used. In our case a more sophisticated process had to be developped so as to extract the magnetic signal from the background and the different sources of fluctuations. For precise MR mesaurement including noise characterization a specific amplification setup had to be designed. was designed. We have designed a longitudinal MOKE imaging setup including a vectorial low noise MR measurement system is 
described in section II. The image processing techniques that were developed so as to enhance the magnetic contrast are presented in section III. Even if used in a longitudinal MOKE configuration, these techniques can be applied to any MOKE configurations. The low noise preamplificer as well as the low frequency noise power supply used to bias the devices under test are presented in section IV. 
The characteristics of the vector magnet that was built are detailed in section V. In section VI, images of magnetization reversal processes as well as magnetic hysteresis loops in active regions of  Planar Hall effect sensors patterned in LSMO oxide thin films are provided. They enable to confirm an expected uniaxial anisotropy. Angle dependent MR measurements are shown to be in accordance with such an uniaxial anisotropy, the magnetic anisotropy constant of the film is then deduced from the magnetic torque calculated with the MR data.

\section{MOKE imaging Setup}
 In our case, the MOKE setup is in longitudinal configuration. It 
is fixed on a vibration free optical table. 
A monochrome OSRAM OSTAR projection power LED source was selected rather than a laser diode since first it avoids usual  interference patterns that superimpose on images when using laser diodes and second it provides a illumination power similar to the laser diode. The blue OSRAM OSTAR LED (ref LE B P1W-EZFZ-24) was selected since the Kerr Effect in LSMO films is larger for the $\lambda$ = 495 nm wavelength compared to other visible wavelength. The incident and reflected beams are perpendicular to each other. They are inclined at 45$^o$ with normal to sample surface. The beam size on the sample is adjusted thanks to a slit placed behind the LED and a converging lens. A polarizer is inserted between the focusing lens and the sample so as to set a p-polarised incident beam. The s-polarized component of the reflected beam is detected by an analyser that is inserted between the sample and the zooming tube directly mounted on a 12 bits Hamamatsu CMOS camera C11440. This reflected s-polarized component provides information of the in plane $M_{x}$ magnetization component through the square of the r$_s$$_p$ reflection coefficient.  The analyser is tilted by an angle equal to $\sqrt{\frac{T_{N} }{{T_{//}}}}$ from the perfect crossed position with respect to the polarizer \cite{Kirchner}, where T$_N$ and T$/s$$/p$ are the transmission coefficient of the polarizer in the normal and parallel direction respectively. This slight tilt provides the optimum contrast in perfect conditions, assuming the illumination is perfectly constant and the camera has no fluctuations \cite{Flament} . 
For this setup we designed a 360$^o$ rotational sample stage. This stage comprises a non-metallic sample holder which can accommodate 5 mm x 10 mm sample and which is fitted to a goniometer which helps to align the sample and chose the orientation of the device to test. 
 
A LABVIEW code was developed to acquire images, perform MR measurements at the same time and control the magnetic field produced by the coils.

The images acquired and stored as numeric array are then processed using a Python script to provide images of magnetization reversal process as well as 
local magnetic hysteresis loop. 

\section{Image processing}

The acquisition time is adjusted so that gray levels of regions of interest are close to the mid range of the 12 bits camera, ie around 2000. Considering that Poisson noise is dominating, any gray level of regions of interest has a standard deviation of about 45. The Kerr effect that occurs when light is reflecting on LSMO films with thiskness ranging from 20 to 60 nm is small. So that, the difference total in gray levels between two opposite saturated state of magnetic area in region of interest is about 100. The signal to noise ratio is thus small, close to 2. In addition to this, the illumination is not spatially homogeneous, the illumination intensity provided by the led suffers fluctuations of up to a few percent, the camera gain may slightly fluctuates, all these fluctuations together are thus of similar amplitudes to the magneto optical signal to measure. As a result, even if image accumulation is performed in order to improve the signal to noise ratio, considering one image as reference and subtract any other raw to this reference image was in our case not an efficient method to get contrasted magnetic images. A specific image processing scheme had to be develop so as to enhance the contrast the images and get a high enough signal to noise ratio in order to deduce from the images the local hysteresis of the magnetic areas in the region of interest. 
Any gray level of pixel (x,y) is proportional the light intensity and to the reflectivity (that a includes the kerr effect) at these coordinates. The incident light intensity at position (x,y) can be expressed by : \\
I(x,y) = $\gamma(x,y)\times I_o \times \beta(t)$\\
where I$_o$ is a constant proportionnal to the averaged LED power, $\gamma$(x,y) corresponds to the normalized  spatial distribution of the light intensity, $\beta$(t) takes into accounts the fluctuations of light intensity and camera gain with time. Values of $\beta$(t) are close to 1 and vary up to a few percent between two acquisition time t$_1$ and t$_2$. The gray level available at pixel (x,y) at the acquired time t can be expressed by : 
$N(x,y) = \alpha(x,y) \times I(x,y) $ \\
where $\alpha(x,y)$  is related to the r$_{sp}$ reflection coefficient index at (x,y) and the relevant magnetic information is related only to the $\alpha$(x,y) values. The image processing scheme must then provide images where pixels values are only related to only $\alpha$(x,y) and not to I(x,y). The full image processing scheme consists in four steps steps : 1- Image accumulation, 2- compensation of illumination fluctuations and any other fluctuations that modify the averaged gray level value between two acquisitions of the same physical state, 3- cancellation of illumination inhomogeneity and modified differential imaging, 4- modification of histogram. These four steps are detailed below.

\begin{enumerate}

\item Image accumulation

Due to the Poisson process of incoming photons, any gray level $N_{ij}$ of pixel has a standard deviation equal to $\sqrt{N_{ij}}$, the signal to noise ratio is described as: $S$/$N$=$N_{ij}$/$\sqrt{N_{ij}}$ = $\sqrt{N_{ij}}$. As a result of the limited dynamic range of the camera, the only way to increase $N_{ij}$ above the full range is to accumulate images and then calculate the average of all acquired images in a given magnetic state. The signal will thus be multiplied by the number of images say $n_{acq}$, while the noise will be multiplied by $\sqrt{n_{acq}}$. As a result, the signal to noise ratio is improved by a factor $\sqrt{n_{acq}}$ as followed:
\begin{eqnarray}
\frac{Signal}{Noise} = \frac{N_{ij} \cdot n_{acq}}{\sqrt{N_{ij} \cdot n_{acq}}} = \sqrt{n_{acq}} \cdot \frac{N_{ij}}{\sqrt{N_ij}} = \sqrt{n_{acq}} \cdot \frac{S}{N}
\end{eqnarray}

\item fluctuations compensation of LED source and camera gain

In non magnetic areas like the substrate or the gold pads the averaged gray levels are not supposed to change when external magnetic conditions are modified. Any changes in these specific areas are thus only due to illumination and camera gain instabilities.  These changes can thus be identified in these specific areas and then finally removed from the whole image using the following process. Let $\bar{N}_1$ be the average gray level of one of these areas for the first image and $\bar{N}_i$ the gray level for any subsequent $i^{th}$ image. Due to fluctuations $\bar{N}_i$ = $\beta_i$ $\bar{N}_1$, where $\beta_i$ is close to 1 and is a measure of the change in camera gain and light intensity between the acquisitions of the first image and the $i^{th}$ one. By dividing the gray levels of all the pixels of the $i^{th}$ image by $\beta_i$, the averages values of the selected area is kept constant, ie the effect of the changes in light intensity and camera gain have been removed in these non-magnetic areas but they have also been removed from the entire image as well. In addition, the relative values between each pixel of any image is preserved by this simple rescaling.

\item cancellation of illumination inhomogeneity and modified differential imaging 

 We first neglect the Poisson noise related to the Poisson process of incoming photons and suppose that there is no fluctuations of illumination nor in camera gain. Assuming, a uniform non magnetic sample, the image of such sample gives the distribution of the illumination. If we divide any subsequent image of this sample by the first image of this sample, we obtain an image where all gray levels are equal to one. Spatial inhomogeneity is thus removed. In the real case of a non uniform sample, let's take a first image, named as ImageSat, as a reference image, when the sample is for example set in a magnetic saturated state. In any next image, named as Imagei, taken in a different physical state, in magnetic material areas the magnetization distribution and then $N_i(x,y)$ may change while in nonmagnetic areas no change will occurr. Let's call $N_{sat}(x,y)$ the gray levels corresponding to ImageSat and $N_i(x,y)$ the gray levels of image$_i$. 
We have : 
\\
$N_{sat}(x,y) = \alpha_{sat}(x,y) \cdot I(x,y)$

$N_i(x,y) = \alpha_i(x,y) \cdot I(x,y) $

where I(x,y) is the illumination distribution and $\alpha_{sat}(x,y)$ and $\alpha_i(x,y)$ the reflection coefficients at the (x,y) position. The difference between $\alpha_i(x,y)$ and $\alpha_{sat}(x,y)$ are small so that \\
$\alpha_i(x,y) = \alpha_{sat}(x,y)  +  \Delta \alpha_i(x,y)\  $

with $ \Delta \alpha_i \ll \alpha_{sat} (x,y) $. 

Instead of subtracting images, if we divided Imagei by ImageSat and multiply the result by a scalar $\lambda$ , for instance equal to half the full range of the camera, the resulting image gets rid of illumination inhomogeneity. Gray levels are all centered around $\lambda$ and the magnetic information is converted in gray levels that are a few percent apart from $\lambda$ value, as expressed in the following equation :
\begin{eqnarray}
\frac{Image2}{ImageSat} \times \lambda =\frac{N(x,y)}{N_{sat}(x,y)} . \lambda\\
 =\left(1 +\frac{\Delta \alpha(x,y)}{\alpha_{sat}(x,y)}\right). \lambda \
\end{eqnarray} 

with $\lambda = \frac{Full \ Dynamic \ Range}{2} = \frac{2^n}{2} $\\

The image resulting from this process provides only the magnetic information $\Delta \alpha(x,y)$ and does not longer depend on the illumination distribution I(x,y). In order to enhance the contrast of the final image a gray level rescaling centered to $\lambda$ is finally required since the gray levels values are centered around $\lambda$  $\pm$ a few \% of the full dynamic range. 


\item modification of histogram  

 The histogram of the images produced by the previous actions is centered around $\lambda$. This histogram is limited to a [N$_{min}$, N$_{max}$] range centered to $\lambda$ with a $(N_{max}-N_{min})$ range much smaller than the full range of the camera (around $\pm$ 400 gray levels compared to the 4096 dynamic range of our 12 bits camera). This histogram is modified in a linear way so that new so that $N_{min}$ is converted to 0 and $N_{max}$ to $2^n$. Any gray level N of the original image is converted in a $N_{new}$ value according to the following equation:\\


$N_{new}=\frac{(2^n}{N_{max}-N_{min})} \cdot \left(N -N_{min}\right) $ \

\end{enumerate}

\section{MR Setup}
Magnetoresistive sensors patterned on epitaxial La$_{0.67}$Sr$_{0.33}$MnO$_3$ (LSMO) thin film have been characterized with this setup. Such samples show a low noticable value of low-frequency noise and thus require a well-suited amplifying system. \cite{guillet, enger2022, enger2023key}.


The magnetoresistive samples were biased with a Yokogawa GS200 DC Voltage/Current Source and a low-noise amplifier based on the AD8421 instrumentation amplifier was deigned for MR characterization.  Among commercially available instrumentation amplifier, as can be deduced from figures 2 and 3, AD8421 is the most suitable choice in order to provide low levels for both low frequency noise and Johnson noise is the resistance range of the MR samples which is typically of a few k$\Omega$. 


	\begin{figure}[t]
\centering
		\includegraphics[width=8cm]{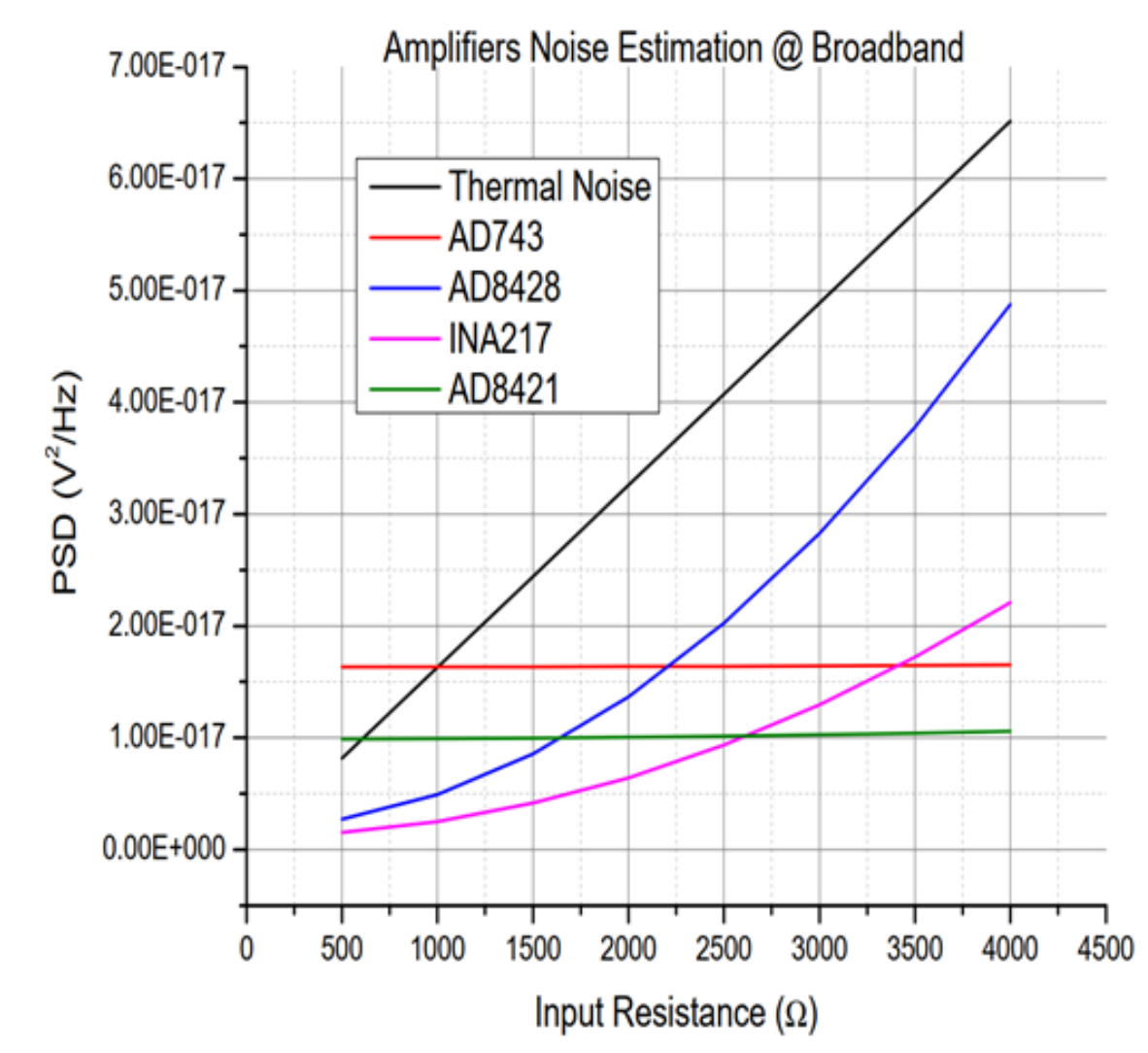}
	\caption{(Color online)  Comparison of amplifiers noise in Broadband.}
	\label{fig:Fig2}
\end{figure}

	\begin{figure}[t]
\centering
		\includegraphics[width=0.48\textwidth]{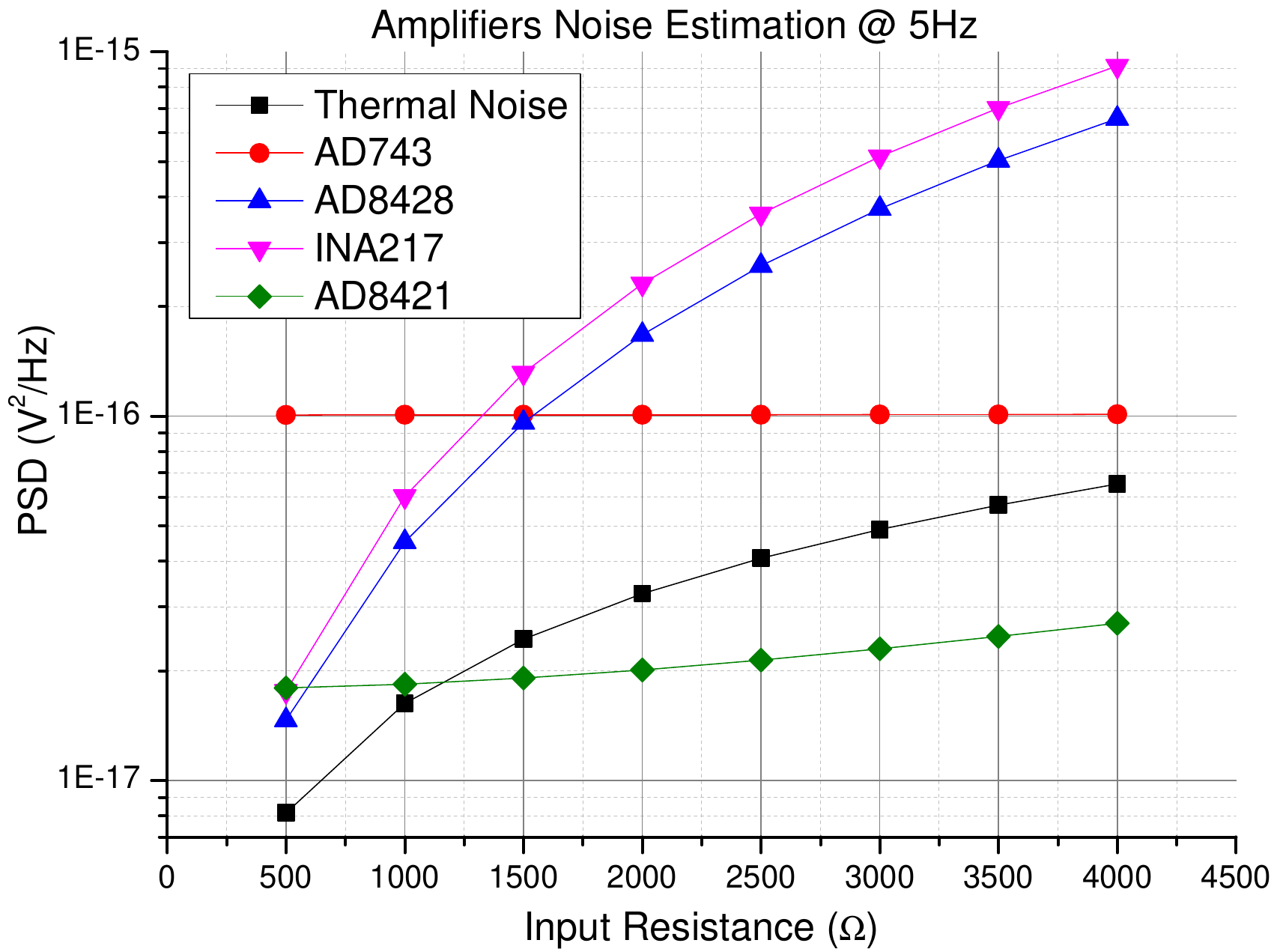}
	\caption{(Color online)  Comparison of amplifiers noise at 5 KHz.}
	\label{fig:Fig2noise.pdf}
\end{figure}

\section{Vector Magnet}

In order to study the magnetic anisotropy in detail, it is useful to change the direction of the applied magnetic field with respect to the sample position.\cite{jye}. In our case we designed a vector magnet that allows to keep the sample in a fixed position for imaging while the field is rotated from 0$^{\circ}$ to 360$^{\circ}$ for angle dependent magneto-resistance measurement.


The home made vector magnet consists of four as identical as possible coils. Two coils are aligned along an x axis and separated by a 40mm gap while two other coils are aligned along an orthogonal z axis and separated by the same 40mm gap. The coils diameter is XXmm, each coils have 450 turns of copper wires whose diameter is 30$\mu$m and a soft ferrite core (ref 0R-49925-IC  from Magnetics supplier) is inserted inside each coil. 
The two sets of two coils are connected to two identical but independent power supplies so as to be able to produce independantly any x and z value of magnetic field.The schematic of coil arrangement and vector component of field is shown in Fig. 3.



	\begin{figure}[t]
\centering
		\includegraphics[width=8cm]{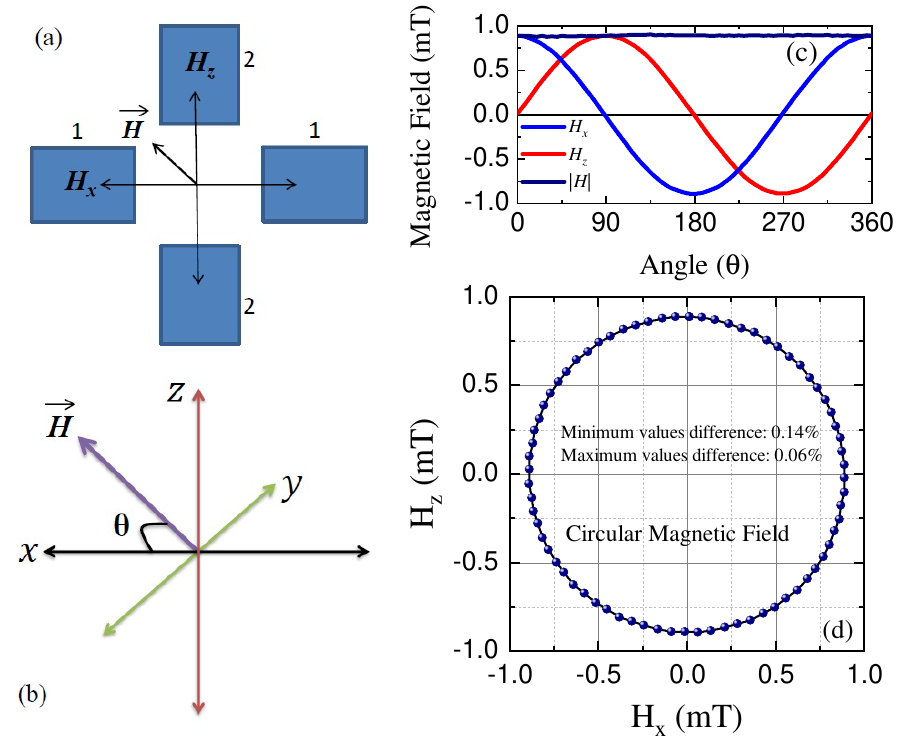}
	\caption{(Color online)  (a-c) Vector magnet Coils arrangement, field vector representation and H$_x$, H$_y$ components.  (d) Polar representation of the field value produced from 0$^{\circ}$ to 360$^{\circ}$}.
	\label{fig:Fig3}
\end{figure}

Even if we tried to make coils identical to each other yet it is not feasible to do so, therefore the coils need to be calibrated separately. For this, 
the $x$ and $z$-components of the field, respectively named $H_x$ and $H_z$, are measured using a gauss meter as a function of the respective current flowing in x aligned and z aligned coils. Once the calibration done, for a given direction of the field in the X-Z plane and a given field value in this direction, one just needs to set appropriate currents according to the following equations : 
%

\begin{eqnarray}
\phi_H = atan(\frac{H_z}{H_x})
\end{eqnarray}

\begin{eqnarray}
H = \sqrt{(H_x)^2 + (H_z)^2}
\end{eqnarray}

The magnitude of rotational magnetic field (H) is also shown in Fig. 3c, it is clearly seen the constant magnitude over 360$^o$ rotation with $<1$\% error.
%

\section{Measurements and results}

%
%
%
%
%
%
%
%

To check the data quality and accuracy of the setup we have performed measurements on LSMO based device. The device is patterned from a LSMO thin film grown on STO(001) 8$^o$ vicinal substrate. In this case , the magnetization is in plane without any out of plane component \cite{Perna2010}. The LSMO thin film is grown using Pulsed Deposition method at substrate temperature of 730 C$^o$ and O$_2$ pressure of 0.02 mbar. The thickness of grown film is measured to be 60 nm. The device is patterned as a Wheatstone bridge as proposed by Henriksen et al \cite{henriksen}, thus eliminating any common mode fluctuations. Each arm of the device is 100 $\mu$m wide and 500 $\mu$m long. Fig. 5 shows the schematics of patterned thin film, easy and hard axis and magnetic field direction. For magneto-resistance electrical measurement gold pads are developed on the device terminals and connections to these gold pads are made using wire bonding.

\begin{figure}
\centering
		\includegraphics[width=8cm]{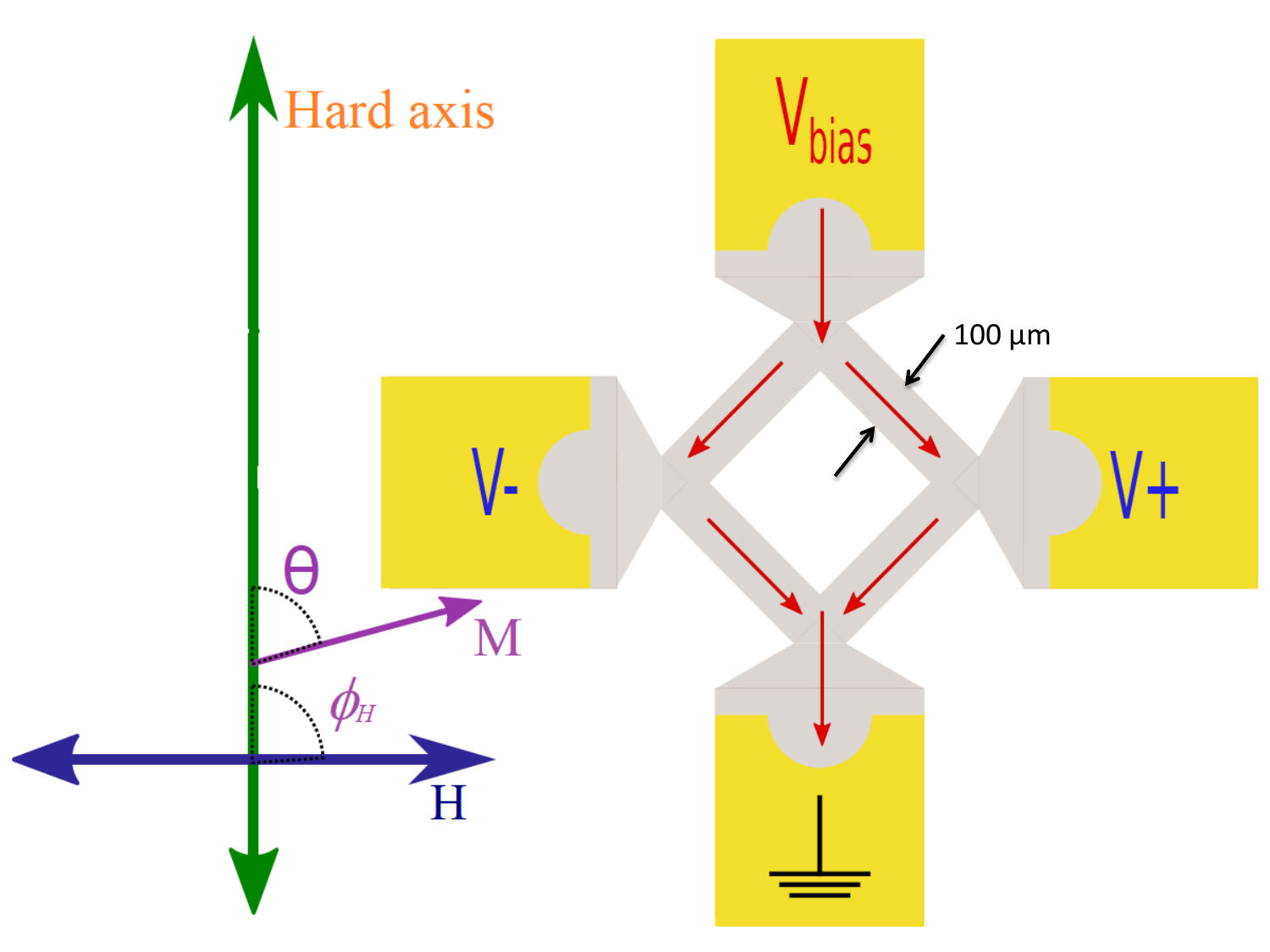}
	\caption{(Color online) Schematic of thin film device representing easy axis, hard axis, magnetic field direction and biasing geometry.}
	\label{fig:Fig5}
\end{figure}

For MOKE characterization, sample were loaded on a sample holder and aligned with fine adjustments using a goniometer so that light beam reflected from sample points to the camera. The images of the sample was captured, processed and stored. The first image is captured at the negative maximum applied field for reference. Then the  magnetic field is swept in step by step mode. At each step many images of the understudy sample are captured and averaged to better the signal to noise ratio. This is done for full cycle of field from negative maximum field to zero to positive maximum and them again zero and back to negative maximum. Finally the captures images are processed as explained in section III. Local MOKE hysteresis loop are then extracted from the porcessed images.

\begin{figure}[h]
\centering
		\includegraphics[width=8cm]{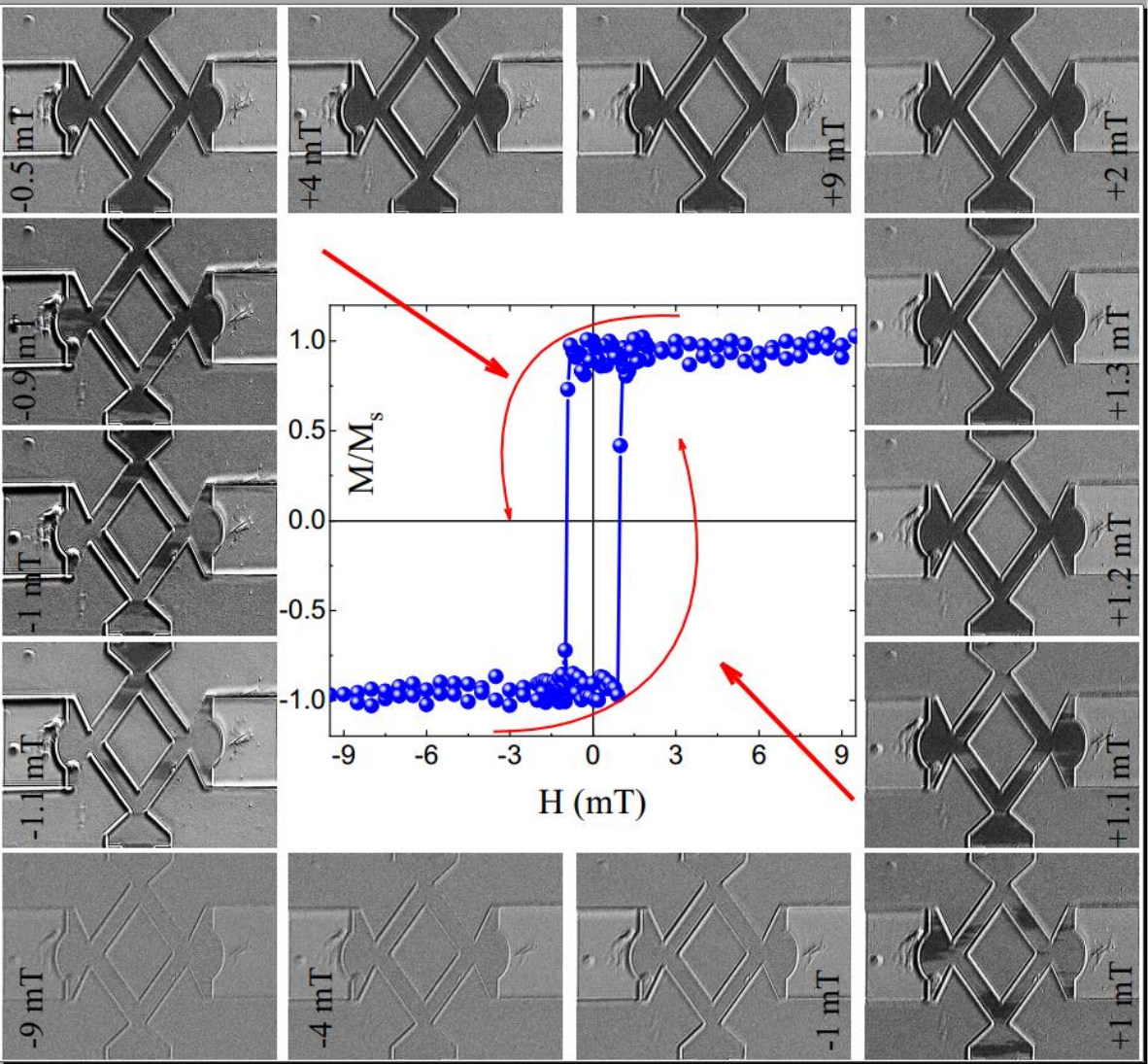}
	\caption{(Color online)  MOKE images and extracted hysteresis loop under applied magnetic field along easy axis at 300 K on LSMO Wheatstone bridge.}
	\label{fig:Fig6}
\end{figure}

\begin{figure}[h]
	\centering
		\includegraphics[width=8cm]{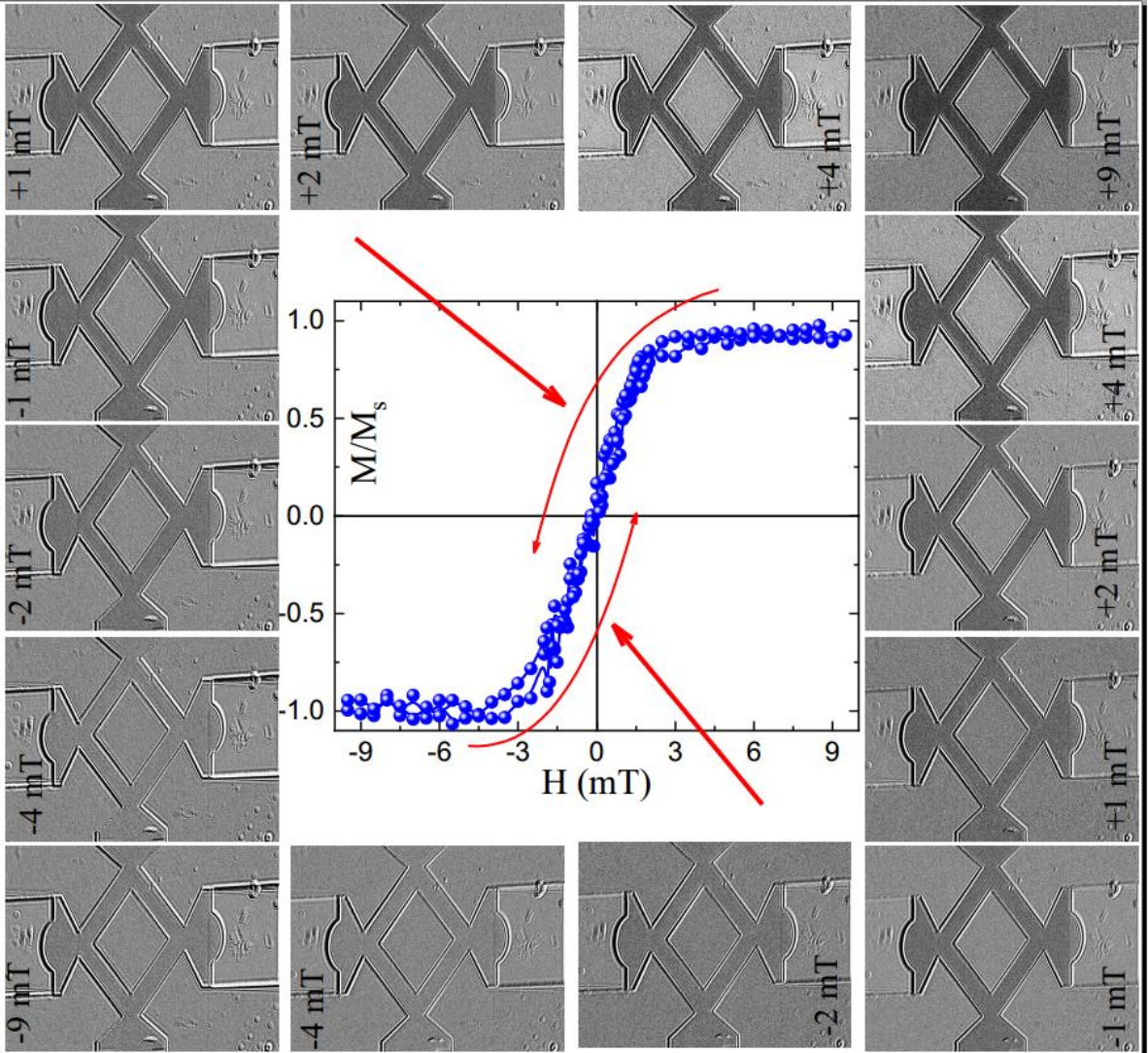}
	\caption{(Color online) MOKE images and extracted hysteresis loop under applied magnetic field along hard axis at 300 K on LSMO Wheatstone bridge.}
	\label{fig:Fig7}
\end{figure}

\begin{figure}[t]
	\centering
		\includegraphics[width=8cm]{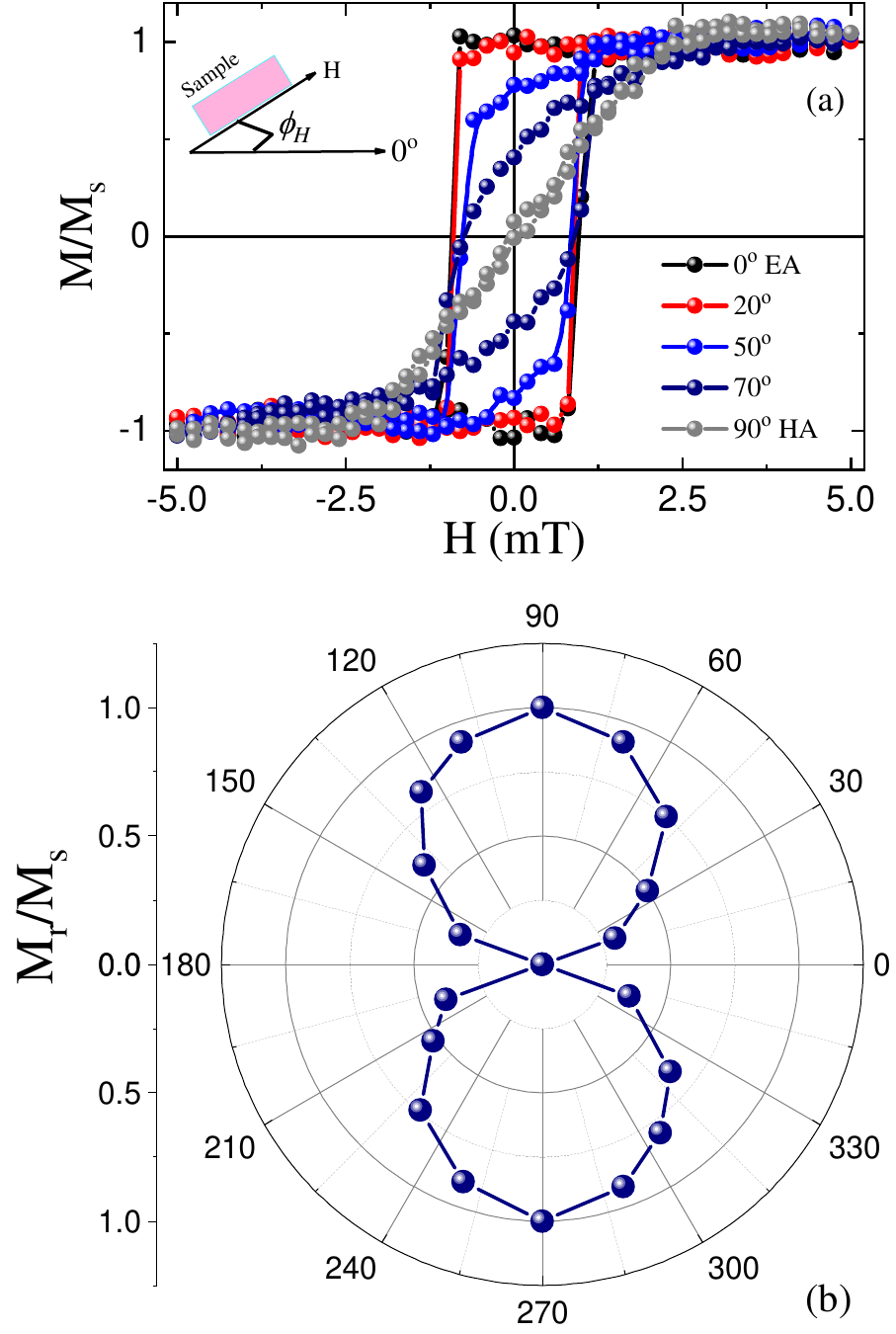}
	\caption{(Color online) (a) Hysteresis loop in measured at different orientation of field with sample. (b) Angle dependent normalized remanent magnetization.}
	\label{fig:Fig8}
\end{figure}


\begin{figure*}[t]
\centering
		\includegraphics[width=16cm]{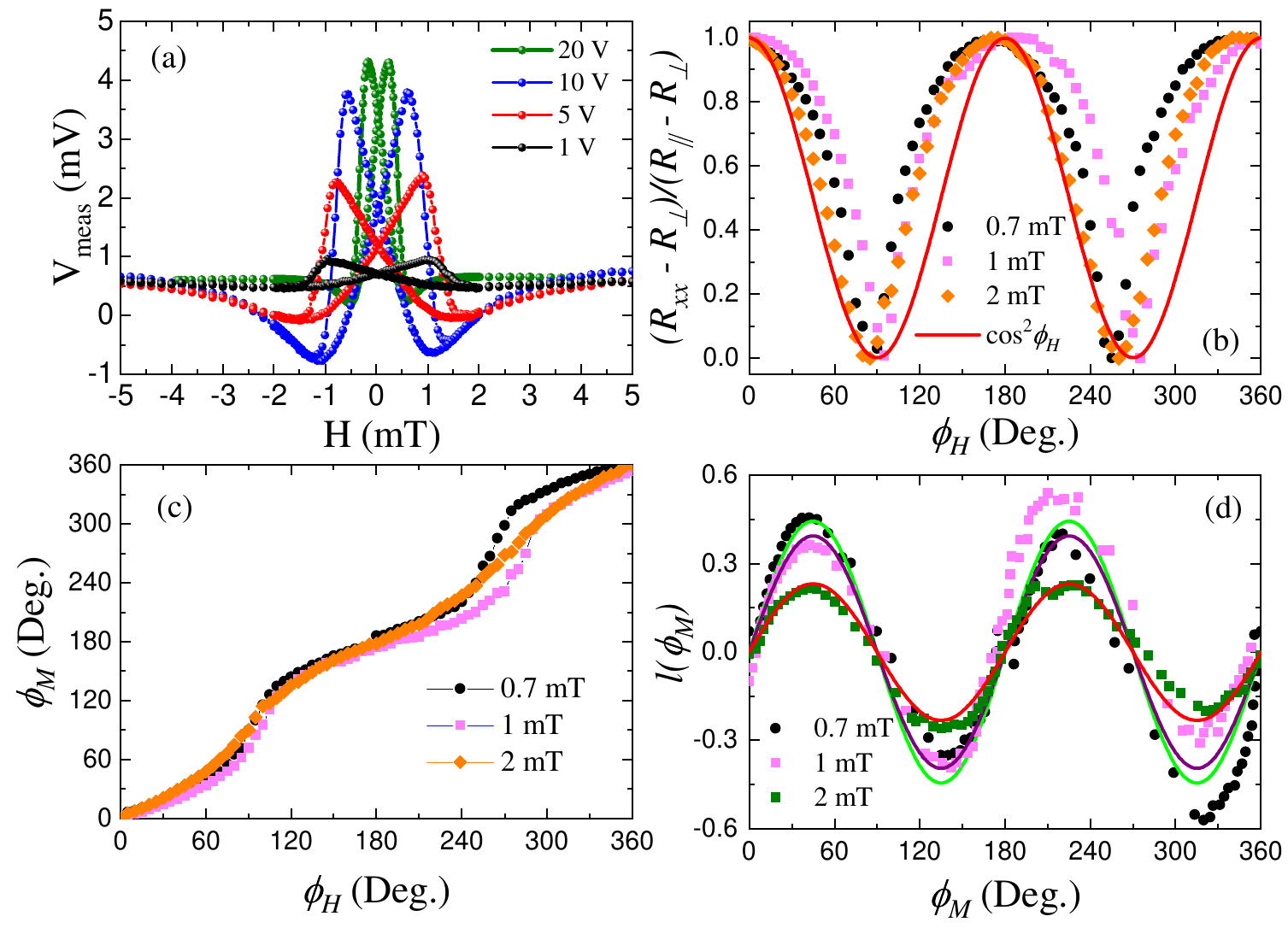}
	\caption{(Color online) (a) Magneto-resistance measured under different biased voltage for LSMO sensor. Inset of (a) is sensitivity. (b) Angular dependency of anisotropic magneto-resistance (AMR) measure at different applied magnetic fields for LSMO sample, red solid line is cos$^2$$\phi_H$ curve. (c) The correlation between $\phi_H$ and $\phi_M$ at different applied magnetic fields. (d) The normalized magnetic torque curve plotted for different applied fields, the solid lines are fitting due to Eq.10.}
	\label{fig:Fig10}
\end{figure*}

Fig. 6 shows the MOKE images of the sample when magnetic field is applied along the easy axis in longitudinal configuration. It can be clearly seen from the Fig. 6  with increasing magnetic field no contrast have been seen in the images since the spins are not magnetized and are randomly orientated in the sample. When the external magnetic field is enough strong the small dark spot appears in the image around +1 mT which indicates nucleation of domains begins at various points in the sample. With further increase in magnetic field it is seen that the domains starts to grow and finally at +1.5 mT magnetic field all the domain oriented in the direction of field and bring the sample to saturation. however we have further increased the to +9 mT to check the contrast level of images, but it was exactly the same for maximum applied as it was for saturation field. To check the magnetization reversal the external applied field is decreased in steps and we found the sample remains magnetized till -1.5 mT of applied field. With further negative field the domains started to shrink and reversed can be seen in Fig. 6. with further increasing negative field the image of the sample looses all the contrast which confirms the magnetization reversal in the thin film. These images are them used to extract the hysteresis loop shown inside the Fig. 6 the clear hysteresis is seen with coercive field as high as 1.5 mT for this sample. We further capture the images of the sample with applied field along hard axis. It is obvious from the Fig. 7 that the magnetization process is rather continuous where no domain nucleation appears at any spot on the sample rather contrast of magnetic area of the sample turns black in the image with increasing application of field. The maximum contrast is seen above -1.5 mT and on reversing the field the spin follow the magnetic orientation. The images are them used to extract MOKE loop placed in side the Fig. 7 clearly indicates no hysteresis in the loop for hard axis and saturation field is around 1.5 mT.

We have plotted the representative angle dependent MOKE loops the sample in Fig. 8a. It is obvious from the figure that the the easy axis shows a large hysteresis whereas the hard axis no such feature is observed, which confirms the reliability of this setup for imaging and MOKE loop tracing. To further check the system we have took MOKE images and extracted MOKR loop at different orientation of the sample with external magnetic field. The orientation dependent MOKE loops are shown in Fig.8a. It is obvious from the figure from easy axis to hard axis orientation the loop  gradually looses the hysteresis it is what is expected in this case. these results confirms the the setup has reasonable accuracy and it is resalable for measurements. Further, we have plotted the normalized remnant magnetization as as function of field orientation of sample in terms of polar plot as shown in Fig. 8b. The shape of plot shows sample have uni-axial magnetic anisotropy. 


Magnetoresistance measurement is another feature added to this setup. In order to characterize anisotropic magnetoresistance (AMR) base low field magnetic sensors for sensitivity region MR measurements are necessary. We have measured the magnetoresistance of the Wheatstone bridge under different biasing voltages. The biasing voltage is applied to the opposite terminals of the bridge and signal is measured from other two terminals as shown in Fig. (ABC) schematic of Wheatstone bridge. Fig. 10a shows the MR measure for the bridge under applied bias voltage of 0 to 20 V. with increasing bias voltage the signal increases and anisotropic field $H_a$ decreases. similar kind of results have been observed when sample is tested on commercially available setup from Lakeshore. Inset Fig 10a the MR data is plotted to fetch the sensitivity detection region. We found that the sample can be useful when biased with 10 V and in the field range of $\pm$0.7 mT. These results are smooth and reasonably good. This measurement shows that the setup can simultaneously perform MOKE and magnetoresistance measurements.

The vector magnet and rotational sample stage enable us to measure the angle dependent magneto-resistance under various constraints like different biasing voltage and different applied magnetic field. Anisotropic magnetoresistance data can be used to study the magnetic anisotropy by determining the magnetic anisotropic constants. Notable point here is that to ensure the rotation of single domain and eliminating the contribution from ordinary MR the applied magnetic field must be larger than saturated field. The anisotropy magneto-resistance can be expressed as:\cite{ye, ahmad, wen}
\begin{eqnarray}
R_{xx} = R_{\perp} +(R_{\parallel} - R_{\perp})cos^2{\phi_M}
\end{eqnarray}
where $\phi_M$ is the angle between magnetization and the current direction I, R$_{\parallel}$ and R$_{\perp}$ is the resistance at $\phi_M$ = 0$^o$ and 90$^o$ respectively.
Fig. 10b shows the angle dependent MR measured under different applied magnetic fields. It is evident from the plot that the AMR curves oscillates between maximum R$_{\parallel}$ and minimum R$_{\perp}$ values. The presence of magnetic anisotropy in the sample the magnetic moments doesnot follow the external magnetic field, thus $\phi_M$ lags behind $\phi_H$ i.e. $\phi_M$<$\phi_H$. Therefore the AMR curves no longer follow cos$^2$$\phi_H$ as evident in the Fig. 10b, the experimental data and cos$^2$$\phi_H$ are not overlapping. The value of $\phi_M$ can be obtained using following relation:\cite{ye, ahmad, wen}
\begin{eqnarray}
\phi_M = acos\left(\sqrt{\frac{R_{xx} - R_{\perp}}{R_{\parallel} - R_{\perp}}}\right)
\end{eqnarray}
The correlation of $\phi_M$ and $\phi_H$ as obtained from Fig. 10b is plotted in Fig. 10c. Using the difference in the $\phi_H$ and $\phi_M$ we can calculate the magnetic torque ($L(\phi_M)$). To compare the magnetic torque curves in different applied fields, normalized magnetic torque was introduced given by the following relation:\cite{ye, ahmad, wen}
\begin{eqnarray}
l(\phi_M) = \frac{L(\phi_M)}{\mu_0 M_s H} = sin(\phi_H - \phi_M)
\end{eqnarray}
where $M_s$ is saturation magnetization, $H$ the applied magnetic field and $\mu_0$ is the permeability constant.
 The torque curves are shown in Fig. 10d in different applied fields. The magnetic torque decreases with increasing applied field and for higher field the torque curve shows smooth behavior which is due to small hysteresis a higher fields.\cite{wen} Further, for uni-axial anisotropy the energy per unit volume of a single domain can be expressed as:\cite{ye, ahmad, wen}
\begin{eqnarray}
E = -\mu_0 M_s H cos(\phi_H - \phi_M) + K_u sin^2 \phi_M 
\end{eqnarray}
where $K_u$ is the uni-axial magnetic anisotropic constant. In the equilibrium state, the angle $\phi_M$ can be calculated by minimizing the Eq. 9 with respect to $\phi_M$ which can gives torque. In other words, the torque acting magnetic moments $M$ due to external magnetic field $H$ is equal in magnitude to the torque due to the magnetic anisotropies of the sample. Thus we can write the normalized torque as following:\cite{ye, ahmad, wen}
\begin{eqnarray}
l(\phi_M) = \frac{K_u\,sin(2\phi_M)}{\mu_0 M_s H}
\end{eqnarray}
The above equation is used to fit the experimental data in Fig. 10d, solid red lines are the fitting. From the fitting parameters we have calculated the value of  $K_u = 89 \pm 2 J/T$ 

The obtained value for LSMO is in the range reported in literature.\cite{gao} The data obtained from this setup is of good quality and confirms the reliability of this setup.

\section{Summary}

We have successfully designed and fabricated custom setup for Magneto Optic Kerr Effect and magneto-resistance measurement. We have tested the setup on LSMO thin film based magnetic sensor. The MOKE images shows nucleation of magnetic domains and magnetization reversal in the thin film strips. The magneto-resistance and angle dependent magneto-resistance data of sensor is obtained successfully. The developed setup is useful for the characterization of magnetic thin films for studying magnetization reversal, domain motion, anisotropy and magneto-resistance.

\section{Acknowledgment}
We acknowledge L. Sylvain and G. Julien for technical support. This project has received funding from the European Union’s (EU) Horizon 2020 research and innovation program under grant agreement No 737116. We acknowledge ByAxon health tech and EU.

\bibliography{References}

\end{document}